\documentclass[twoside,a4paper,11pt]{sca}
\usepackage{graphicx}
\usepackage{amssymb,amsmath}
\usepackage{hyperref}
\usepackage{movie15}
\usepackage{natbib}  
\begin{document}
\pagenumbering{arabic}
\pagestyle{myheadings}
\thispagestyle{empty}
{\flushright\includegraphics[width=\textwidth,bb=90 650 520 700]{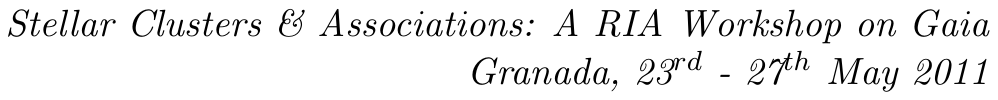}}
\vspace*{0.2cm}
\begin{flushleft}
{\bf {\LARGE
%
Origin of Extended Star Clusters 
%
}\\
\vspace*{1cm}
Narae Hwang$^{1}$
%
}\\
\vspace*{0.5cm}
%
$^{1}$
National Astronomical Observatory of Japan, 
Tokyo 181-8588, Japan\\
%
\end{flushleft}
%
\markboth{
Origin of Extended Star Clusters 
}{ 
%
Narae Hwang 
}
\thispagestyle{empty}
\vspace*{0.4cm}
\begin{minipage}[l]{0.09\textwidth}
\
\end{minipage}
\begin{minipage}[r]{0.9\textwidth}
\vspace{1cm}
\section*{Abstract}{\small
We have discovered new extended star clusters (ESCs) in a nearby dIrr galaxy
NGC 6822. These clusters are the nearest sample of ESCs available to date. The
key characteristic of ESCs is their large size compared to typical globular
clusters even though the two cluster populations are rather similar in terms of
other parameters, i.e., color and luminosity. Several scenarios have been
suggested to explain the formation of ESCs. However, the currently known ESCs
may be a mixture of populations with heterogeneous formation histories. Future
observational and theoretical studies are expected to better constrain the
origins of ESCs as well as to increase their sample size.
\normalsize}
\end{minipage}
%
%
\section{Observational Characteristics of ESCs\label{intro}}

ESCs are one of the new cluster populations being
discovered with the advance of observational studies of star clusters.
They are characterized by a relatively large size, that is, the half-light
radii R$_{\rm h} \gtrapprox 10$ pc, compared to typical globular clusters (GC)
with R$_{\rm h} \sim 2$ -- $3$ pc. Some examples may include faint fuzzy
clusters discovered in SB0 type galaxies \citep{bro02,hwa06} and extended
clusters found in the halo of M31 \citep{hux05,hux11}. Even in the Milky Way
galaxy, there are rather extended GCs found in the outer halo \citep{vdb04}.
The ESCs in dwarf galaxies had not been known until the discovery of new ESCs
in an isolated dIrr galaxy NGC 6822 \citep{hwa05,hwa11}. Those new ESCs in NGC
6822 are the nearest sample of ESCs available to date (D$\approx 500$ kpc).

Another noteworthy characteristic of ESCs is their spatial distribution. It
is shown that the ESCs in NGC 6822 are preferentially aligned along the old
stellar halo that lies almost perpendicular to the HI disk like structure (see
Fig.1 in \citealt{hwa11}). Faint fuzzy clusters in NGC 5195 and M51 also
exhibit elongated spatial distribution, which is not followed by red compact
clusters \citep{hwa06,hwa08}. This elongated distribution of ESCs is different
from the one reported for faint fuzzy clusters in NGC 1023 that are usually
found in the galaxy disk \citep{lar00}. For ESCs in M31, no distinct spatial
distribution is apparent. However, it is suggested that many ESCs in M31
appear to be associated with stellar streams in the halo \citep{col09,hux11}.

\section{Scenarios at the moment}

Figure 1 shows the correlations between $R_{\rm h}$ and $M_{\rm V}$ of various
types of objects including GCs, LMC, UCDs (ultra compact dwarfs), DGTOs (dwarf globular transition objects), UFDs (ultra faint dwarfs), dSph galaxies,
as well as ESCs in M31, M33, and NGC 6822.
There are at least two important points to note. Firstly, ESCs are found in
almost every type of galaxies ranging from dwarfs (e.g., NGC 6822, Scl-dE1) to giant
spirals (e.g., M31) and ellipticals (e.g., NGC 5128). Secondly, there are two subclasses of
ESCs with different luminosity: one with $M_{\rm V} <-9.4$ and the other with
$M_{\rm V} >-8.0$. The bright ESCs are found to occupy the common parameter
space with UCDs or DGTOs, while the faint ESCs extend toward UFDs.

Various scenarios for the origins of those ESCs have been proposed and some
reviews on each scenario and exemplary ESCs are given in \citet{hwa11}. A
brief summary is as follows: (1) Remnants or cores of tidally stripped dwarf
galaxies may form ESCs. (2) Collisions of two or more star clusters in star
cluster complexes or super-star cluster may make ESCs. (3) Star clusters are
born in various sizes and some ESCs may survive the disruption under the weak
tidal field. (4) Another scenario very recently pointed out by \citet{gie11} is that slow expansion of initially compact clusters under the tidal
field may have produced ESCs whose size evolution can be explained by adopting a
line of constant relaxation time in the mass-radius diagram.

Usually, bright ESCs could be the remnant of disrupted dwarf galaxies
or the multiple star cluster collisions, while faint ESCs may have evolved
under the optimal tidal field to survive or to expand to their current forms.
But no observational evidence is available yet to justify such dichotomy and/or to prefer any scenario over the others.
However, every scenario involves the tidal interactions and/or mergers on the
scale of star clusters or galaxies that should be the major physical
drivers that make ESCs.

For the better understanding and clarification of the ESCs and their possible
use for the galaxy evolution studies, future observations with improved
photometric depth and spectral resolution as well as wider spatial coverage are
required. In view of this expectation, GAIA will play a crucial role to survey
even more ESCs in the Milky Way and to reveal their true nature and origin.

\begin{figure}[t]
\center
\includegraphics[scale=0.53,angle=-90]{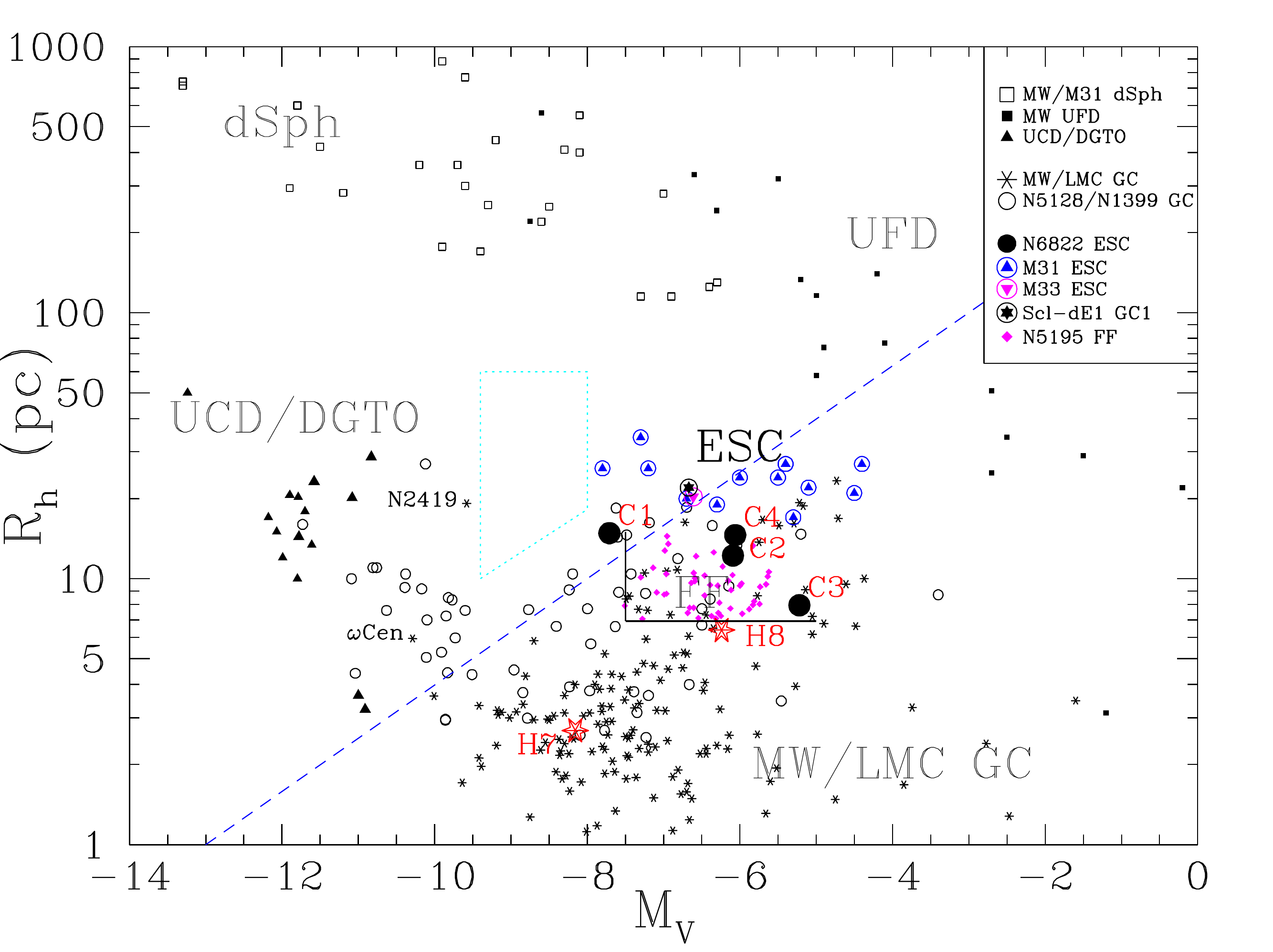}
\caption{\label{fig1} $M_V$ vs. half light radii ($R_h$) diagram of GCs
including the Galactic GCs and LMC GCs \citep{vdb04}, ESCs including M31 ESCs
\citep{hux05,hux11}, M33 ESC \citep{sto08}, Scl-dE1 ESC \citep{dac09} as well as
NGC 6822 ESCs \citep{hwa11}. Bright and faint ESCs are separated by a box in dotted lines at the center. Please refer to \citet{hwa11} for more details.}
\end{figure}

\small  
%
\section*{Acknowledgments}   
%
I acknowledge invaluable collaborations with Myung Gyoon Lee, Hong Soo Park,
Won-Kee Park, and Sang Chul Kim in the course of the study on ESCs in NGC 6822.
I also thank Mark Gieles for bringing my attention to the cluster relaxation
and expansion for the ESC formation.


\bibliographystyle{aa}
\bibliography{mnemonic,n6822_nhwang_arxiv}

\end{document}